# A Taxonomy of Single-Turn Textual Prompt Patterns

Vennila Sooben*[1] and Eugene Syriani†[1]

[1]DIRO, Université de Montréal, Canada

29 June 2026

**Abstract**

Large language models (LLMs) are now widely employed in software development and everyday use. Interacting with LLMs requires crafting prompts, which range from simple ad hoc sentences to extensive, detailed, and structured instructions. Knowledge about prompt engineering has been documented in several surveys and catalogs in the literature. However, the term "prompt pattern" is defined differently across sources, and existing works have classified prompt patterns in different ways. In this report, we present a taxonomy of prompt patterns for single-turn, text-based interactions. Following a reproducible method, we identified 30 unique and canonical prompt patterns, organized along two dimensions.

## 1 Introduction

Large language models (LLMs) are increasingly used in software development and in everyday contexts. In software engineering, they are used to support tasks such as code generation, code completion, debugging, test generation, documentation, code review, requirements analysis, and code explanation [1]. They are also used for traditional natural language processing tasks (e.g., writing, summarization), and other tasks (e.g., information seeking and planning) [2]. Across these contexts, users interact with LLMs primarily by formulating prompts.

This growing reliance on prompts has motivated interest in prompt engineering: the design of prompt inputs to guide LLM behavior [3]. However, prompting knowledge is difficult to reuse when it is expressed only through isolated examples or informal advice. This makes it difficult to compare prompts, reuse prompting knowledge across contexts, and report prompt designs systematically in empirical studies involving LLMs.

Prompt patterns address this issue by capturing recurring textual structures that can be adapted across tasks and domains. This is particularly useful in a field where users with different backgrounds and levels of expertise interact with LLMs. Patterns can provide a common vocabulary, clarify which components may be included in a prompt, and support decisions about when a particular prompting structure is appropriate. However, existing work uses different terminology, scopes, and classification criteria when describing prompt patterns and related prompting techniques [4, 5, 6, 7, 8]. As a result, the available knowledge remains difficult to consolidate into a stable and reusable catalog.

In this report, **we present a taxonomy of prompt patterns for single-turn, text-based interactions with LLMs**. We focus on reusable textual structures that can be instantiated within a single prompt. Using a reproducible classification process, we identify 30 unique and canonical prompt patterns from existing literature. We organize them primarily by prompting strategy and annotate each pattern with the prompt template components in which it can be instantiated. The resulting taxonomy supports the description and reuse of prompt patterns, particularly in software engineering research involving LLMs. The taxonomy is available online[1] to support reuse and future extension.

*vennila.sooben@umontreal.ca

†syriani@iro.umontreal.ca

In Section 2, we specify the scope of the taxonomy and define key terminology. In Section 3, we discuss related work and, specifically, the literature on which we have constructed our taxonomy. In Section 4, we describe the method we followed to design the taxonomy of prompt patterns. In Section 5, we present the elements to define a prompt pattern and present a few examples. In Section 6, we discuss the limitations of the taxonomy and possible future uses. Finally, we conclude in Section 7.

# 2 Scope of prompt patterns

In this section, we define the scope of the prompts and patterns we consider for the taxonomy.

## 2.1 Prompts

A prompt is the input through which a user specifies a task, provides context, and guides the expected response of an LLM. Prompts can range from simple ad hoc sentences to extensive, detailed, and structured instructions. In text-based interactions, a prompt may include a task description, domain context, examples, constraints, role specifications, or indications about the expected output format [9]. As a result, prompting goes beyond asking a question. It requires deciding which information should be included, how the prompt should be organized, and how explicitly the expected behavior of the model should be described.

The quality and usefulness of LLM outputs depend in part on the formulation of the prompt. A vague prompt can lead to an incomplete, overly general, or poorly structured answer, whereas a more explicit prompt can guide the model toward a more relevant and usable response. For example, the prompt "`Review this code`" gives the model little information about the expected type of review. In contrast, a structured prompt, like the one in Listing 1, can specify the programming language, the review objective, the kinds of issues to identify, and the desired output format. This example shows how prompt structure can make the expected response more explicit. Prompt engineering has emerged as a way of improving LLM behavior by designing prompt inputs [2]. It is the process of selecting, organizing, and formulating textual instructions given to an LLM.

```
As a senior developer, review the following {{programming language}} code:          role, task
{{specific code}}                                                                input, context
Consider:                                                                           constraints
  1. Code quality and adherence to best practices
  2. Potential bugs or edge cases
  3. Performance optimizations
  4. Readability and maintainability
  5. Any security concerns
Suggest improvements and explain your reasoning for each suggestion.                  directive
Classify the suggestions into 3 categories: CRITICAL, IMPORTANT, and NICE TO HAVE.  output format
```

Listing 1: Example of a structured prompt for code review

## 2.2 Definitions

We first define a few related terms used in this report.

[1] https://geodes.iro.umontreal.ca/promptspec-catalog/

**Definition 1.** A *prompt* or *prompt instance* is a string submitted to an LLM without modification in a given interaction.

**Definition 2.** A *prompt template* is a reusable prompt structure containing placeholders or variable parts that can be filled or adapted for a specific task, input, or context before submission as a prompt instance.

For example, Listing 1 is a prompt template because it contains placeholders, like `{{paste your code}}`. A prompt instance is the string after all placeholders or variable parts have been resolved.

**Definition 3.** A *prompt workflow* is an interaction-level procedure composed of multiple steps, such as a sequence of prompt instances, model responses, user revisions, or tool interactions.

We focus on prompt patterns that can be instantiated within a single prompt, as opposed to interaction-level prompt workflows. The taxonomy we propose focuses on single-turn, text-based interactions with LLMs. Single-turn prompts are a basic unit of interaction and can be studied independently from the dynamics of multi-turn conversations with an LLM. Multi-turn interactions often involve sequences of prompts, intermediate model responses, user feedback, revisions, and context updates. These elements introduce additional design concerns that are outside the scope of this taxonomy.

Text remains a common medium for specifying tasks, constraints, examples, and output expectations. Although multimodal prompting is important, a taxonomy of textual prompt patterns provides a bounded starting point for consolidating reusable patterns before considering patterns that involve images, audio, video, or other modalities.

Prompt engineering knowledge is often expressed through examples, guidelines, or informal recommendations. While these forms are useful, they can be difficult to reuse systematically because they are often tied to specific tasks or domains. Prompt patterns address this issue by abstracting recurring textual structures that can be instantiated in different prompt instances.

**Definition 4.** A *prompt pattern* is a recurring and reusable textual structure, expressed in a canonical form, that can be instantiated within a prompt to support a specific intention in an interaction with an LLM.

It is a canonical structure that can be adapted to different scenarios and combined with other structures inside a concrete prompt instance. To be considered a pattern, such a structure should be recurring, associated with a recognizable prompting intention, and general enough to allow multiple concrete instantiations. For example, each component labeled on the right in Listing 1 illustrate prompt chunks that may instantiate different prompt patterns.

**Definition 5.** A *prompting strategy* is a category that groups prompt patterns according to the primary prompting intention they support.

For example, it can provide examples, like few-shots prompting [10], or guide reasoning, like chain-of-thoughts prompting [4].

**Definition 6.** A *prompt component* is a functionally identifiable part of a prompt that contributes to specifying the interaction with an LLM.

Mao et al. [9] identify the following seven prompt components. Listing 1 illustrates some of them.

- *Profile/Role* specifies who or what the model is expected to act as.
- *Directive* specifies the main task, instruction, or question given to the model.
- *Context* provides background information or input content needed to perform the task.

- *Workflows*[2] specify steps or procedures that the model should follow within a single prompt.
- *Output Format/Style* specifies the expected structure, format, or style of the response.
- *Constraints* specify restrictions, requirements, or conditions that the response should satisfy.
- *Examples* provide sample inputs, outputs, or demonstrations of the expected behavior.

Prompt components are not necessarily mandatory, mutually exclusive, or fixed in position. A short prompt may consist only of a directive, whereas a structured prompt may combine several components.

Therefore, the taxonomy excludes prompting practices that cannot be expressed as reusable textual structures within a single prompt. These include multi-turn conversational workflows, agentic workflows, tool-use protocols, retrieval-augmented generation pipelines, fine-tuning methods, and other model adaptation techniques. Such techniques may interact with prompt patterns, but they are not themselves prompt patterns under Definition 4.

# 3 Related work

We position our taxonomy with respect to existing work on prompt engineering surveys, prompt pattern catalogs, and prompt structuring frameworks.

## 3.1 Prompt engineering surveys and catalogs

Several surveys and catalogs have organized prompt engineering knowledge for LLMs. The five works discussed in this subsection constitute the initial corpus for our taxonomy. In this section, we use the term *prompting technique* to refer to the broader terminology used in existing surveys. It includes prompt patterns, but also broader methods such as multi-step workflows, tool-use techniques, retrieval-based methods, multimodal prompting, or model adaptation approaches.

White et al. [4] introduce a catalog of prompt patterns intended to support prompt engineering with conversational LLMs. They explicitly frame prompting knowledge using the notion of reusable patterns and draw an analogy with software design patterns [11]. Their catalog is particularly relevant to our work because it treats prompt patterns as reusable solutions to recurring problems in LLM interaction.

Schulhoff et al. [5] provide a broad survey of prompting techniques and propose a structured vocabulary and taxonomy for prompt engineering. Their work covers a wide range of text-based prompting techniques, as well as prompting techniques for other modalities. It is relevant to our taxonomy because it consolidates a large set of prompting concepts that are often discussed separately in the literature.

Sahoo et al. [6] survey prompt engineering techniques and applications for LLMs and vision-language models. They organize prompting approaches by application area. Their survey is useful for identifying prompting practices across different application contexts, although its scope is broader than reusable textual techniques for single-turn prompts.

Vatsal and Dubey [7] survey prompt engineering methods for different natural language processing tasks. They organize prompting methods according to the tasks in which they have been used and discuss their reported performance across datasets and models. Their work provides a task-oriented view of prompting techniques, which complements catalogs organized by pattern or technique type.

Fagbohun et al. [8] provide an empirical categorization of prompting techniques for large language models from a practitioner-oriented perspective. They classify prompting techniques into seven categories.

[2] A prompt workflow is distinct from the *Workflow* prompt component, which refers to steps or procedures specified inside a single prompt.

Their classification includes prompting techniques at a broader level of granularity than the prompt patterns considered in this report.

These five sources cover prompt patterns, prompting techniques, task-oriented prompting methods, application-oriented surveys, and practitioner-oriented categorizations. However, they differ in terminology, scope, granularity, and classification criteria. This motivates the need to consolidate their reported concepts into a taxonomy of reusable prompt patterns for single-turn, text-based interactions.

### 3.2 Prompt structuring frameworks

Recent work has also proposed frameworks for structuring prompts and describing prompt properties. Prompt Canvas proposes a practitioner-oriented framework that synthesizes prompt engineering knowledge into a visual canvas [12]. Its goal is to provide a practical resource for learning and applying prompt engineering techniques. TELeR proposes a taxonomy of prompt properties organized around Turn, Expression, Level of Details, and Role, with the goal of supporting more comparable benchmarking of complex tasks [13]. These frameworks show the need for more systematic ways of describing prompts and prompt design decisions.

Other frameworks define component-based structures for constructing prompts. AUTOMAT organizes prompt writing around elements such as the model role, user persona or audience, target action, output definition, mode or style, atypical cases, and topic whitelisting [14]. CO-STAR structures prompts around Context, Objective, Style, Tone, Audience, and Response [15]. COSTAR-A extends this structure by adding an Answer component and evaluates the resulting framework for point-of-view questions [16]. RCFOR structures prompts around Role, Context, Focus, Objective, and Response [17]. These complement Mao et al.'s [9] seven prompt components we described in Section 2.2.

Prompt-design guidelines provide another form of structuring support. For example, Bsharat et al. [18] propose 26 principles for prompt and instruction design and evaluate their effects across several LLMs. Such guidelines do not necessarily define prompt components or prompt patterns, but they provide practical advice for making prompts more explicit and effective.

Our work is complementary to these frameworks and guidelines. Prompt Canvas provides a broad practitioner-oriented view of prompt engineering, whereas TELeR characterizes prompts according to properties relevant to benchmarking. Frameworks such as AUTOMAT, CO-STAR, and RCFOR provide component-based structures for constructing individual prompts. In contrast, our taxonomy focuses on consolidating recurring textual structures into a deduplicated and scope-bounded catalog of prompt patterns. Rather than describing complete prompts through general properties, checklists, or guidelines, we identify canonical reusable structures that can be instantiated as all or part of a single-turn, text-based prompt.

### 3.3 Challenges for a taxonomy

Existing work provides substantial knowledge about prompt engineering, but this knowledge remains difficult to reuse as a catalog of prompt patterns. We identify four limitations that motivate our taxonomy.

First, existing sources do not use the term *prompt pattern* consistently. Some works use it to refer to reusable textual structures, while others use related terms to include broader prompting techniques, multi-turn workflows, or methods that depend on external tools or model adaptation. Similar concepts may also appear under different names across sources, while the same name may refer to different concepts. This terminological variation makes it difficult to determine whether two reported techniques are equivalent, overlapping, complementary, or distinct.

Second, existing catalogs often combine concepts at different levels of granularity. A single catalog may include small prompt components, complete prompt templates, reasoning strategies, interaction workflows, retrieval mechanisms, or fine-tuning approaches. This makes it difficult to isolate the reusable textual structures that can be instantiated within a single prompt.

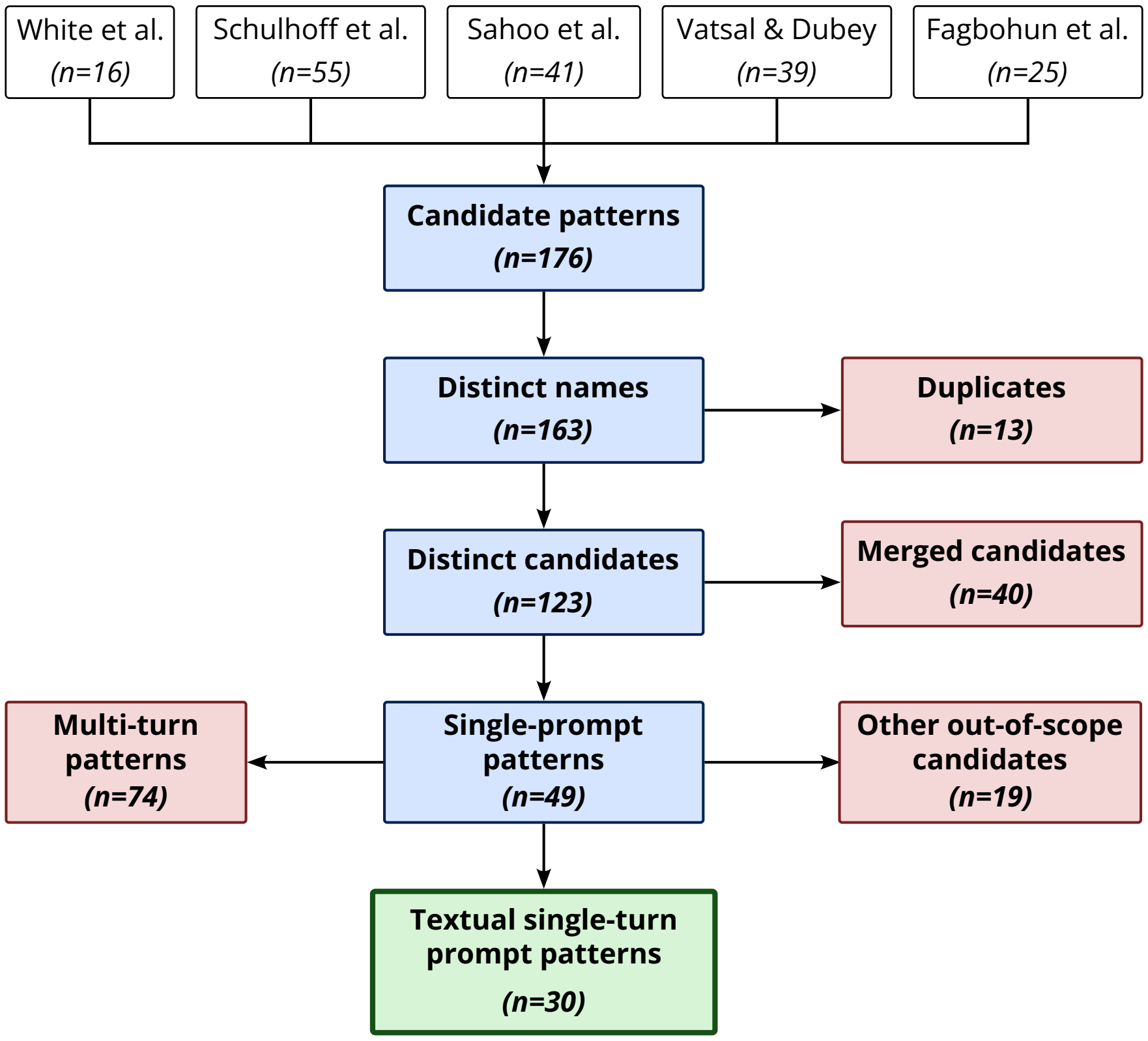

Figure 1: Process to extract textual single-turn prompt patterns

Third, prompt engineering concepts are represented in heterogeneous ways. Some concepts are described through prose, others through examples, and others through pseudo-templates or named techniques. This heterogeneity makes it difficult to compare concepts systematically and to identify when several descriptions correspond to the same underlying pattern.

Fourth, the procedures used to collect, deduplicate, and classify patterns are not always explicit. This limits reproducibility and makes it difficult to extend existing catalogs as new prompting practices emerge. Because prompt engineering is still evolving, a reusable taxonomy should make its inclusion, exclusion, deduplication, and classification decisions inspectable.

For a taxonomy of prompt patterns to support reuse, the construction process must therefore make explicit which source concepts are retained, excluded, merged, or represented as variants of a more general pattern.

# 4 Method

Our objective is to derive a taxonomy of reusable prompt patterns from existing prompt engineering literature. The taxonomy focuses on textual structures that can be instantiated within single-turn prompts, as defined in Section 2. We followed a reduction process that extracts candidate patterns from selected sources, removes duplicates, merges overlapping concepts, and filters out concepts that do not satisfy the scope of the taxonomy.

Figure 1 summarizes the process. We use the term *candidate pattern* to refer to any named prompting concept extracted from the selected sources. At this stage, a candidate pattern is not necessarily a prompt pattern under Definition 4. After deduplication, semantic consolidation, and scope filtering, we obtain the final set of 30 prompt patterns for single-turn textual prompts.

## 4.1 Source selection and candidate extraction

We selected five sources that organize prompt engineering knowledge through prompt patterns, prompting techniques, task-oriented methods, application-oriented surveys, or practitioner-oriented categorizations [4, 5, 6, 7, 8]. These sources were selected because they provide named prompting concepts that can be extracted and evaluated according to our definition of prompt pattern.

From each source, we extracted the concepts presented as prompt patterns, prompting techniques, prompting methods, or prompting strategies. For each candidate pattern, we recorded its original name, source, description, available examples or templates, and any classification provided by the source.

The extraction produced 176 candidate patterns: 16 from [4], 55 from [5], 41 from [6], 39 from [7], and 25 from [8]. For [5], we considered only text-based prompting techniques from their dataset, which resulted in 55 extracted candidates rather than the full set of 58 techniques.

## 4.2 Name deduplication

The extracted candidates contained exact and near-exact duplicate names across sources. We first normalized these duplicates to obtain a set of distinct names. For example, concepts reported as *Chain-of-Thought* and *CoT* were treated as the same named concept when their descriptions referred to the same prompting idea. This step removed 13 duplicates and reduced the set to 163 distinct names.

## 4.3 Semantic consolidation

We then consolidated candidates that described the same underlying prompting intention using different names or slightly different formulations. For example, *Persona* [4], *Role* [5], and *Multi-Personas* [8]. all refer to the general idea of assigning a specific role, perspective, or expertise to the model. In such cases, we selected one canonical name and recorded the others as aliases.

We also identified candidates that were better represented as variants or combinations of broader patterns rather than as independent canonical patterns. For example, *Contrastive CoT Prompting* combines example-based prompting with the elicitation of intermediate reasoning [19]. Rather than treating it as fully independent, we represented it as a variant or combination of broader canonical patterns when appropriate.

This semantic consolidation step merged 40 candidates and reduced the set to 123 distinct candidate patterns.

## 4.4 Scope filtering

We then applied the scope defined in Section 2. The goal was to retain only reusable textual patterns that can be instantiated within a single prompt.

First, we excluded candidate patterns that apply to multi-turn prompts. These candidates describe interaction-level procedures involving a sequence of prompts, intermediate model responses, user feedback, tool calls, or iterative revisions. For example, a technique such as *Cumulative Reasoning*, where reasoning steps are generated, evaluated, and accumulated until a stopping condition is reached [20], describes a multi-step interaction procedure rather than a structure instantiated within one prompt. This step excluded 74 candidates and reduced to 49 single-turn candidate prompt patterns.

Second, we excluded 19 additional candidates that did not satisfy the definition of textual prompt pattern:

- **Preprocessing techniques** ($n$ = 4): techniques that describe operations performed before prompt construction rather than textual structures instantiated within a prompt. For example, *Code Prompting* rewrites the prompt as programming code [21].

- **Prompt attributes** ($n = 3$): dimensions that can modify how a prompt pattern is written, but that are not standalone reusable structures. For example, *Style Prompting* can specify the writing style, tone, mood, pacing, genre, or level of detail of a prompt [22].
- **Prompting examples** ($n = 8$): concepts that correspond to specific examples, tasks, or use cases without defining a reusable canonical structure. For example, *Grammar Correction* is a specific task rather than a pattern [8].
- **Non-textual or training-based concepts** ($n = 4$): concepts that require non-textual inputs, model training, fine-tuning, or other model-level interventions. For example, *Show-me versus Tell-me Prompting* requests the LLM to output a graphical representation rather than a textual explanation [8].

After this final filtering step, 30 candidates remained. These candidates constitute the final set of textual single-prompt patterns in the taxonomy.

## 4.5 Taxonomy construction

We organize the 30 prompt patterns in two complementary ways. First, each pattern is assigned to a two-level taxonomy based on its primary prompting intention. Second, each pattern is annotated with the prompt component type(s) in which it can be instantiated.

### 4.5.1 Prompting strategy categories

The primary organization of the taxonomy is based on prompting strategies (see Definition 5). We constructed five top-level categories:

- **In-Context Learning**: patterns that use, omit, or structure examples to guide the model's response.
- **Reasoning**: patterns that elicit, structure, decompose, or separate intermediate reasoning before producing an answer.
- **Output Control**: patterns that constrain the form, structure, schema, procedure, or verification of the generated output.
- **Context Control**: patterns that assign a role, perspective, or contextual grounding to guide the model's response.
- **Meta-Directives**: patterns that operate on the prompt, query, or interaction itself, such as improving, refining, or reframing the request.

These categories were synthesized from recurring distinctions observed across the five sources. For example, *Context Control* is borrowed from White et al. [4]. We also retain *In-Context Learning* and *Reasoning* as top-level categories because these families recur across prompt engineering surveys [5, 6]. We use *Output Control* (originally termed *Output Customization* in [4]) to group patterns that constrain the structure, format, or verification of the response. Finally, we introduce *Meta-Directives* to group patterns whose primary intention is to act on the prompt, query, or interaction itself rather than directly on the domain task.

Each top-level category is further divided into at least two subcategories. Figure 2 shows the resulting category structure. Below, we define the 14 subcategories. Those with a * are subcategories we introduced or renamed to harmonize related concepts across sources.

- **Zero-Shot**: patterns that guide the model without providing task-specific examples.

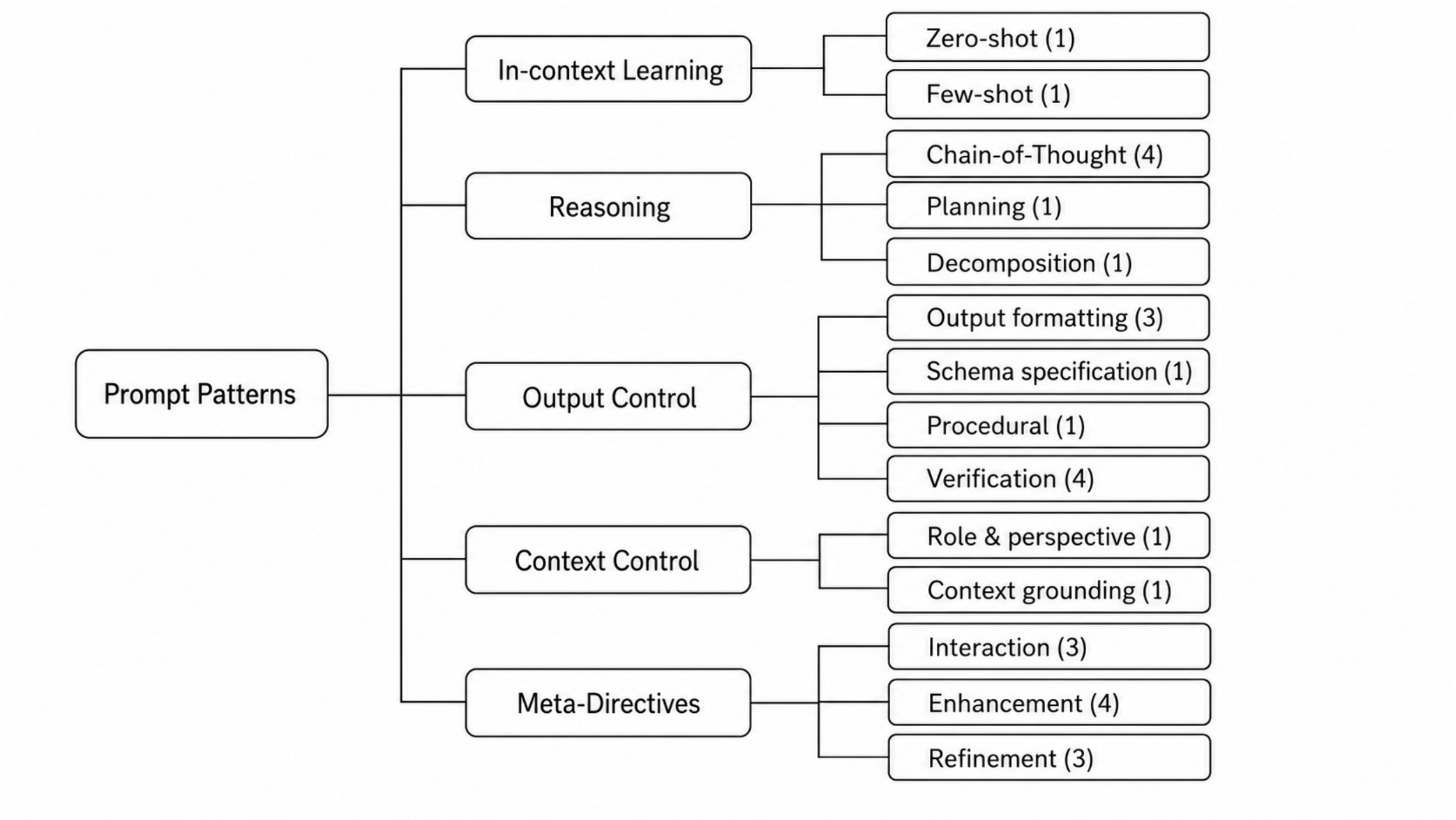

Figure 2: Taxonomy categories and their subcategory. Numbers in parenthesis indicate the number of patterns classified in each subcategory.

- **Few-Shot**: patterns that guide the model through one or more examples.
- **Chain-of-Thought**: patterns that elicit intermediate reasoning before the final answer.
- **Planning***: patterns that separate plan construction from task execution.
- **Decomposition***: patterns that break a task into smaller subproblems to be solved sequentially.
- **Output Formatting***: patterns that specify the expected form, layout, or presentation of the output.
- **Schema Specification***: patterns that define an explicit output schema, such as required fields, types, or length constraints.
- **Procedural**: patterns that specify a procedure or sequence of steps to follow within a single prompt.
- **Verification**: patterns that ask the model to check, validate, or assess the generated answer.
- **Role & Perspective***: patterns that assign a role, identity, expertise, or perspective to the LLM.
- **Context Grounding**: patterns that anchor the response in provided context or external information.
- **Interaction**: patterns that shape how the model should interact with the user or handle the request.
- **Enhancement***: patterns that improve, enrich, or complete a prompt or generated answer.
- **Refinement***: patterns that rewrite, clarify, or transform the query before answering.

Table 1: Coverage of prompt patterns by prompt component.

| **Pattern** | **Profile/Role** | **Directive** | **Context** | **Procedural Steps** | **Examples** | **Output Format/Style** | **Constraints** |
|---|---|---|---|---|---|---|---|
| *In-context Learning* | | | | | | | |
| ZeroShot | | • | | | | | |
| FewShot | | • | | | • | | |
| *Reasoning* | | | | | | | |
| ChainOfThought | | | | • | | | |
| StructuredCoT | | | | • | | • | |
| ComplexCoT | | | | • | | | |
| PlanAndSolve | | | | • | | | |
| LeastToMost | | | | • | | | |
| ReverseCoT | | | | • | | | |
| *Output Control* | | | | | | | |
| VisualizationGenerator | | | | | | • | |
| Recipe | | • | | • | | | |
| SelfVerification | | | | • | | | • |
| SelfCalibration | | | | • | | | • |
| FactCheckList | | | | • | | | • |
| Reflection | | | | • | | | • |
| OutputAutomater | | | | | | • | |
| Template | | • | | | | • | |
| SchemaSpecs | | | | | | • | |
| *Context Control* | | | | | | | |
| Persona | • | | • | | | • | |
| ContextManager | | • | • | • | | | |
| MetaLanguageCreation | | • | • | | | | |
| *Meta-Directives* | | | | | | | |
| FlippedInteraction | | • | | | | | |
| GamePlay | | • | | | | | |
| InfiniteGeneration | | • | | | | | |
| QuestionRefinement | | • | | | | | |
| RAR | | • | | • | | | |
| AlternativeApproaches | | • | | • | | | |
| RE2 | | • | | | | | |
| RefusalBreaker | | • | | | | | • |
| InstructionSelection | | • | | • | | | |
| CognitiveVerifier | | • | | • | | | |
| **Total** | **1** | **16** | **3** | **16** | **1** | **6** | **5** |

### 4.5.2 Component annotation

In addition to the prompting-strategy taxonomy, we annotated each pattern with the prompt component type(s) in which it can be instantiated. These annotations describe the functional location of a pattern within a prompt. Therefore, a pattern may receive more than one component type annotation.

We use the seven component types proposed by Mao et al. [9] (see Definition 6). However, we renamed the *Workflow* component label to *Procedural Steps* to avoid confusion with the notion of prompt workflow defined in Definition 3. This component type refers to steps or procedures specified inside a single prompt, whereas a prompt workflow refers to an interaction-level procedure involving multiple prompts, responses, revisions, or tool interactions. The resulting component annotation is shown in Table 1.

## 5 Prompt patterns

We now describe the prompt patterns retained in the taxonomy. We first define the descriptive schema used to document each pattern, then summarize the 30 patterns in the catalog, and finally provide complete descriptions of five representative patterns. The full catalog is available online.[1]

## 5.1 Pattern description schema

Each prompt pattern is documented using a common description schema. The schema is intended to make pattern descriptions comparable across categories without introducing a formal language for prompt patterns. Each pattern description contains the following fields:

**Name.** A unique canonical name used to refer to the pattern in the taxonomy. We use the notation *Category » Subcategory @ Component(s)* to summarize each pattern classification.

**Classification.** The category and subcategory of the pattern, according to the prompting strategy taxonomy introduced in Section 4.5, together with the prompt component type(s) in which the pattern can be instantiated.

**Intent.** A concise description of the prompting intention supported by the pattern and the motivation for using it.

**Application conditions.** The situations in which the pattern is appropriate, together with cases in which the pattern should be avoided or used with caution.

**Canonical structure.** A generic structure that describes how the pattern can be instantiated. We describe the structure using two parts: *variables*, which identify the variable information needed to instantiate the pattern, and a *template*, which shows how these variables can be arranged in a prompt.

**Consequences.** The expected benefits and limitations of using the pattern.

**Example.** One or more concrete prompt instances illustrating the pattern.

**Variants.** Alternative forms of the pattern, including variants reported in prior work or variants obtained by combining the pattern with other prompt patterns.

**Other names.** Alternative names used in the source literature for the same or closely related pattern.

The canonical structure is not intended as a formal syntax. It provides a compact and reusable way to communicate the recurring textual structure captured by the pattern. Depending on the pattern, it may be instantiated as a complete prompt or as one segment of a larger prompt that combines several patterns.

Table 2 summarizes the 30 prompt patterns in the taxonomy. For each pattern, the table reports its intent, number of variants, and source provenance.

## 5.2 Sample pattern descriptions

We illustrate the description schema with five representative patterns from different categories.

### 5.2.1 Persona

We retain this pattern from White et al. [4]. They present it as a pattern for assigning a point of view, role, or persona to the LLM so that the generated output reflects the kinds of details, style, and output types associated with that persona.

**Name.** *Persona*

**Classification.** *Context Control » Role & Perspective @ Profile/Role*, *Context*, *Output Format/Style*

Table 2: Summary of the 30 prompt patterns in the taxonomy

| Pattern | Short description | Variants | Sources |
|---|---|---|---|
| *In-context Learning* | | | |
| ZeroShot | Instructs the task directly without worked examples. | 3 | [5, 6, 7] |
| FewShot | Supplies worked input-output examples before the target task. | 2 | [5, 6] |
| *Reasoning* | | | |
| ChainOfThought | Elicits explicit intermediate reasoning before the final answer. | 9 | [5, 6, 7, 8] |
| StructuredCoT | Organizes reasoning with explicit program-like control structures. | 2 | [6, 7] |
| ComplexCoT | Generates multiple detailed reasoning paths before answer selection. | 2 | [5, 7] |
| PlanAndSolve | Plans subproblems before executing stepwise solution steps. | 2 | [5, 7] |
| LeastToMost | Decomposes tasks from simpler subproblems to harder ones. | 2 | [5, 7] |
| ReverseCoT | Works backward from the desired output to infer the inputs or steps that produce it. | 1 | [5] |
| *Output Control* | | | |
| VisualizationGenerator | Produces visual or tabular representations of information. | 1 | [4] |
| Recipe | Specifies a reusable ordered procedure for completing tasks. | 1 | [4] |
| SelfVerification | Checks and revises its answer before finalizing output. | 1 | [5] |
| SelfCalibration | Estimates answer confidence and revises or abstains when uncertain. | 1 | [5] |
| FactCheckList | Lists factual claims needed to verify the output. | 1 | [4] |
| Reflection | Critiques an initial answer and improves the response. | 2 | [4, 8] |
| OutputAutomater | Requires a structured, machine-readable output format (e.g., JSON, XML, CSV). | 1 | [4] |
| Template | Fills a reusable template while preserving its structure. | 1 | [4] |
| SchemaSpecs | Constrains output to a specified schema or field structure. | 2 | [5, 8] |
| *Context Control* | | | |
| Persona | Adopts a role or persona to shape responses. | 3 | [4, 5, 8] |
| ContextManager | Controls the contextual scope, included or excluded information, and disciplinary lenses used during response generation. | 2 | [4, 8] |
| MetaLanguageCreation | Defines custom shorthand semantics for subsequent prompt use. | 1 | [4] |
| *Meta-Directives* | | | |
| FlippedInteraction | Makes the model ask questions to reach a goal. | 2 | [4, 8] |
| GamePlay | Frames the task as a rule-governed game. | 1 | [4] |
| InfiniteGeneration | Continues generating repeated outputs until a stop condition. | 1 | [4] |
| QuestionRefinement | Improves or reformulates the user query for clarity. | 1 | [4] |
| RAR | Rephrases and expands the question before answering it. | 2 | [5, 6] |
| AlternativeApproaches | Lists multiple viable approaches to the same task. | 1 | [4] |
| RE2 | Re-reads the question before producing an answer. | 1 | [5] |
| RefusalBreaker | Reframes refused requests into answerable alternative phrasings. | 1 | [4] |
| InstructionSelection | Selects among candidate instructions before executing the task. | 1 | [5] |
| CognitiveVerifier | Generates subquestions whose answers support final verification. | 1 | [4] |

**Intent.** The *Persona* pattern instructs the LLM to adopt a specific role, identity, expertise, or perspective when generating its response. The intent is to shape the response according to the knowledge, style, voice, criteria, or level of detail associated with the specified persona. The pattern is useful when the user knows the type of expert, stakeholder, or entity that should guide the response, but does not know all the output details that such a role would normally consider.

**Application conditions.** This pattern is appropriate when the response should reflect a particular expertise, audience, perspective, or communicative style. It is useful when the persona provides implicit expectations about vocabulary, tone, explanation depth, output type, or evaluation criteria. It can also be used to scope the model's attention toward aspects that are salient for the specified persona. It should be used with caution when the assigned persona may introduce unsupported authority, bias the response toward a narrow viewpoint, or suggest expertise that the model cannot verify. It should not replace explicit task instructions when precise criteria, constraints, or output requirements are needed.

**Canonical structure.** The pattern can be instantiated by specifying the persona to adopt, the question or task to answer, optional aspects to focus on, and optional input data to consider.

*Variables.*

- `persona`: the role, identity, expertise, perspective, or entity assigned to the model.
- `question`: the task, question, or instruction to answer.
- `focus`: optional aspects that the persona should pay particular attention to.
- `input_data`: optional input data or contextual information.

*Template.*

```
Act as {{persona}}.
Provide outputs that {{persona}} would create.
Pay particular attention to {{focus}}.                                optional
{{question}}
{{input_data}}                                                        optional
```

**Consequences.** The pattern can make the response more aligned with the expected role, style, expertise, or audience. It can also make prompts more concise by encoding several expectations through the persona description. However, the persona may bias the answer, hide uncertainty, or produce an impression of authority that is not justified. For simulated entities or systems, the model may also introduce assumptions or synthetic details that are not present in the prompt. The pattern is therefore more reliable when combined with explicit instructions, constraints, output specifications, or verification steps.

**Example.** For example, a user may ask the model to answer as a senior software engineer, security reviewer, software architect, technical writer, requirements analyst, math tutor, or other domain-specific persona.

```
Act as a security reviewer.
Provide outputs that a security reviewer would create.
Pay particular attention to authentication, authorization, input validation, and data
    exposure risks.
Review the following code and identify potential security issues:
{{source_code}}
```

**Variants.** Variants include assigning a professional role, a stakeholder perspective, a domain expert identity, or multiple personas. In a multi-persona variant, the prompt may ask the model to consider a task from several roles before producing a final response. The persona may also be a simulated system or non-human entity, such as a terminal, database, or other interface. In such cases, the prompt should specify how the user input should be interpreted and what type of output the simulated entity should produce. When the simulation requires repeated exchanges, it becomes closer to an interaction workflow and falls outside the single-turn scope of this taxonomy.

**Other names.** Related names include *Role Prompting*, *Role*, and *Multi-Personas*.

### 5.2.2 Context manager

This pattern consolidates two related source concepts. White et al. [4] introduce *Context Manager* as a prompt pattern for specifying, emphasizing, removing, or resetting context in an LLM interaction. Fagbohun et al. [8] describe *Cross-disciplinary Prompting* as a technique for combining knowledge from multiple disciplines to guide responses to interdisciplinary tasks, as employed in [23]. We retain these concepts under *ContextManager* because both explicitly control the contextual frame used by the model: White et al. emphasize inclusion, exclusion, and reset of context, whereas cross-disciplinary prompting specifies the domain perspectives or knowledge frames that should ground the response.

**Name.** *ContextManager*

**Classification.** *Context Control » Context Grounding @ Context*, *Directive*, *Procedural Steps*

**Intent.** The *ContextManager* pattern explicitly controls the contextual frame that the LLM should use when generating its response. The intent is to keep the response grounded in a specified scope, selected information, or disciplinary perspective, while reducing irrelevant, outdated, or unintended contextual influence. The pattern is useful when the prompt contains multiple pieces of information, when only part of the available context is relevant, when some information should be excluded, or when the answer should connect concepts across selected domains.

**Application conditions.** This pattern is appropriate when the user needs to delimit the scope of the response, emphasize relevant information, prevent the model from using unrelated context, or guide the model through a specific contextual lens. It is useful for tasks such as code review, requirements analysis, document analysis, comparison of alternatives, and interdisciplinary explanation. For example, a user may ask the model to explain a technical concept using analogies from another discipline, or to analyze a software artifact only with respect to security concerns. It should be used with caution when the excluded information may be necessary to answer correctly, or when the selected contextual lens may oversimplify the target concept. It should also be used carefully in conversational settings, because instructions to reset or discard context may remove useful prior instructions or constraints.

**Canonical structure.** The pattern can be instantiated by specifying the response scope, the information or perspectives to include, the information to exclude, the question or task to answer, and optional input data.

*Variables.*

- `scope`: the topic, task, document, criterion, discipline, or boundary within which the model should answer.
- `include_items`: the information, concepts, facts, criteria, disciplines, or perspectives that the model should consider.

- `exclude_items`: the information, concepts, facts, criteria, or perspectives that the model should ignore.
- `question`: the task, question, or instruction to answer.
- `input_data`: optional input data or contextual material.

*Template.*

```
Set the scope to {{scope}}.
Consider only the following: {{include_items}}
Ignore the following: {{exclude_items}}
{{question}}
{{input_data}}                                                    optional
```

**Consequences.** The pattern can make the intended context of the response more explicit and reduce context drift. It can improve relevance by directing the model toward specific information, criteria, or domain perspectives. It can also support interdisciplinary explanations by making explicit which disciplines or conceptual frames should be connected. However, overly restrictive context instructions may lead to incomplete answers if relevant information is excluded. Cross-disciplinary framing may also produce superficial analogies or misleading mappings if the relationship between the domains is underspecified. In conversational settings, reset instructions may remove prior instructions or hidden assumptions that the user did not intend to discard. The pattern is therefore more reliable when the included and excluded context are specified explicitly.

**Example.** For example, a user may ask the model to review code only from a security perspective while ignoring formatting or naming issues.

```
Set the scope to security review.
Consider only the following: authentication, authorization, input validation, data
    exposure risks.
Ignore the following: formatting, naming conventions, code style preferences
Review the following code and identify security issues:
{{source_code}}
```

**Variants.** Variants include inclusion-only context control, exclusion-only context control, scoped analysis, context reset, and cross-disciplinary context grounding. An inclusion-only variant specifies what the model should consider without listing exclusions. An exclusion-only variant specifies what the model should ignore while leaving the rest of the context available. A cross-disciplinary variant specifies one or more disciplinary lenses that should be used to explain, analyze, or connect concepts. A reset variant asks the model to discard prior context and restart within a specified scope. The reset variant is mainly relevant in multi-turn conversations; within this taxonomy, it is retained only when expressed as a single-prompt instruction that controls the context used for the current response.

**Other names.** Related source concepts include *Context Manager*, *Context Control*, *Scoped Prompting*, and *Cross-disciplinary Prompting*.

### 5.2.3 Self-calibration

This pattern is derived from the self-evaluation procedure studied in [24] and reported as self-calibration by Schulhoff et al. [5]. In the original procedure, the model first proposes an answer and is then asked to estimate whether that answer is correct. Because our taxonomy focuses on single-turn textual prompts, we retain the single-prompt form in which the model is asked to answer, report its confidence, and revise or abstain when confidence is low.

**Name.** *SelfCalibration*

**Classification.** *Output Control » Verification @ Constraints*, *Procedural Steps*

**Intent.** The *SelfCalibration* pattern instructs the LLM to answer a question and then estimate its confidence that the answer is correct. The intent is to make uncertainty explicit and to support a decision about whether to accept, revise, or abstain from the answer. The pattern is useful when the user wants the model to distinguish between confident and uncertain answers rather than always producing a definitive response.

**Application conditions.** This pattern is appropriate when the task involves uncertainty, incomplete information, factual recall, classification, or decision support. It is useful when users need a confidence signal to decide whether an answer should be accepted, revised, or checked by another source. It should be used with caution because self-reported confidence is not necessarily well calibrated. It should not be used as the only verification mechanism for high-stakes decisions or tasks requiring factual correctness.

**Canonical structure.** The pattern can be instantiated by asking the model to answer the question, report a confidence estimate, justify the estimate briefly, and revise or abstain when the confidence is below a specified threshold.

*Variables.*

- `question`: the question or task to answer.
- `confidence_min`: the minimum confidence value the LLM can output when it reports being completely uncertain.
- `confidence_max`: the maximum confidence value the LLM can output when it reports being completely certain.
- `threshold`: optional confidence threshold below which the model should revise or abstain.

*Template.*

```
Answer the question: {{question}}
Then state your confidence from {{confidence_min}} to {{confidence_max}} that the answer
    is correct, with a brief justification.
If your confidence is below {{threshold}}, revise your answer or state that you are not
    sure.
    optional
```

**Consequences.** The pattern can make uncertainty more visible and may discourage unsupported definitive answers. It can also help users identify answers that require additional checking. However, the confidence estimate is generated by the model itself and may not correspond to a calibrated probability of correctness. The model may also provide a plausible justification for an incorrect confidence estimate. The pattern is therefore more reliable when combined with external verification, source checking, or domain review.

**Example.** For example, a user may ask the model to report confidence for each identified issue before deciding which findings require manual verification.

```
Review the following code for potential security issues:
{{source_code}}
For each issue you report, state your confidence from 0% to 100% that the issue is
    present, with a one-line justification.
If your confidence for an issue is below 60%, revise the issue or state that you are not
    sure.
```

**Variants.** Variants include confidence scoring, confidence thresholds, abstention rules, and revision after low confidence. In a two-step variant, the model first generates an answer and is then prompted again with the question and its proposed answer to estimate whether the answer is correct. This two-step variant becomes an interaction workflow when implemented as separate prompts. In this taxonomy, the retained pattern is the single-prompt form that asks the model to answer and self-calibrate within the same prompt.

**Other names.** No alternative names were retained from the source corpus.

### 5.2.4 ReverseCoT

This pattern is derived from the *RCoT* method introduced by Xue et al. [25] and reported as *Reversing Chain-of-Thought* by Schulhoff et al. [5]. In the original method, the model reconstructs the problem from a generated solution, compares the reconstructed problem with the original problem, and uses detected inconsistencies as feedback for revision. Because our taxonomy focuses on single-turn textual prompts, we retain the form in which reconstruction, comparison, and possible revision are encoded within one prompt.

**Name.** *ReverseCoT*

**Classification.** *Reasoning » Chain-of-Thought @ Procedural Steps*

**Intent.** The *ReverseCoT* pattern instructs the LLM to reason backward from a candidate answer to check whether the answer is consistent with the original question. The intent is to reconstruct the problem, assumptions, or conditions that would lead to the candidate answer, compare this reconstruction with the original problem, and use any inconsistencies as feedback for revision. The pattern is useful when the user wants the model to verify an answer through backward reasoning rather than only produce a forward solution.

**Application conditions.** This pattern is appropriate when a task has an answer that can be checked against the original question, constraints, or input data. It is useful for mathematical reasoning, logical reasoning, program comprehension, debugging, and other tasks where inconsistencies between an answer and the original problem can be identified. It should be used with caution for open-ended or subjective tasks, where there may be no clear reconstruction to compare against the original prompt. It should also be used with caution because the model may rationalize an incorrect answer instead of identifying genuine inconsistencies.

**Canonical structure.** The pattern can be instantiated by asking the model to produce or use a candidate answer, reconstruct the problem or assumptions that would lead to that answer, compare the reconstruction with the original question, and revise the answer if inconsistencies are found.

*Variables.*

- `question`: the question, problem, or task to answer.

- `input_data`: optional input data or contextual information.
- `candidate_answer`: optional candidate answer to check; if absent, the model first generates one.

*Template.*

```
{{question}}
{{input_data}}
First, state a candidate answer directly.
Then reconstruct the problem, assumptions, or conditions that would lead to this answer.
Compare the reconstruction with the original question.
List any inconsistencies.
If inconsistencies are found, revise the answer; otherwise keep the candidate answer.
```

**Consequences.** The pattern can expose mismatches between a generated answer and the original problem. It can also support answer revision by turning detected inconsistencies into feedback. However, the backward reasoning may produce a post hoc rationalization rather than an independent verification. The pattern may also add verbosity and require the user to inspect whether the reconstruction and comparison are meaningful. It is therefore more reliable when the task has explicit constraints and when the comparison step is made concrete.

**Example.** For example, a user may ask the model to first report candidate findings and then reason backward from each finding to the code evidence that would make it valid.

```
Review the following code for potential security issues:
{{source_code}}
First, state the candidate security findings directly.
For each finding, reconstruct the code path, inputs, or assumptions that would make the
    finding valid.
Compare this reconstruction with the original code.
List any inconsistencies between the reconstruction and the code.
If inconsistencies are found, revise or remove the finding; otherwise keep it.
```

**Variants.** Variants include answer-first justification, backward reasoning from a provided answer, and reconstruction-based verification. In the answer-first variant, the model first states an answer and then provides reasoning that supports it. In the reconstruction-based variant, the model reconstructs the problem from the generated answer, compares the reconstructed problem with the original problem, and uses inconsistencies as feedback for revision. When these steps are implemented as separate prompts, the technique becomes an interaction workflow; in this taxonomy, the retained pattern is the single-prompt form that encodes the reconstruction, comparison, and revision steps within one prompt.

**Other names.** Related names include *Reversing Chain-of-Thought* and *RCoT*.

### 5.2.5 Schema specification

This pattern is consolidated from related source concepts that constrain the admissible structure or values of the model output, including output formatting, answer-space Schulhoff et al. [5] describe output formatting and answer-space specification as part of their discussion of answer engineering, where answer space denotes the domain of values that an answer may contain. Fagbohun et al. [8] describe constrained-vocabulary prompting as a technique for restricting the model's response to a predefined vocabulary. We retain these concepts under *SchemaSpecs* because they all constrain the structure, fields, admissible values, or vocabulary of the generated output.

**Name.** *SchemaSpecs*

**Classification.** *Output Control » Schema Specification @ Output Format/Style*

**Intent.** The *SchemaSpecs* pattern instructs the LLM to produce an answer that follows a specified schema, field structure, answer space, or controlled vocabulary. The intent is to reduce variation in the output and make the response easier to parse, compare, validate, or reuse in downstream tasks. The pattern is useful when the user needs the answer to conform to explicit structural or value-level requirements.

**Application conditions.** This pattern is appropriate when the response must follow a predictable structure, such as a JSON object, table, CSV row, XML fragment, form, or list of predefined labels. It is useful for extraction, classification, evaluation, data collection, and other tasks where outputs need to be compared or processed systematically. It is also appropriate when the answer should be restricted to a defined set of values or terms. It should be used with caution for exploratory or open-ended tasks, where a rigid schema may omit relevant information or force uncertain content into predefined fields. It should not be treated as a guarantee that the output will be syntactically valid or semantically correct; external validation may still be required when the output is consumed by software.

**Canonical structure.** The pattern can be instantiated by specifying the question or task, the required schema, the fields to populate, and optional value constraints or vocabulary restrictions.

*Variables.*

- `question`: the task, question, or instruction to answer.
- `schema`: the required output structure, such as JSON, CSV, Markdown, XML, a table, or a custom field structure.
- `fields`: the fields, columns, labels, or structural elements that must be populated.
- `allowed_values`: optional allowed values, labels, terms, or vocabulary items for one or more fields.
- `input_data`: optional input data or contextual information.

*Template.*

```
Answer the following question: {{question}}
Use the following output schema: {{schema}}
Populate the following fields: {{fields}}
Use only the following allowed values or vocabulary when specified: {{allowed_values}}
    optional
{{input_data}}                                                          optional
```

**Consequences.** The pattern can make outputs more consistent, machine-readable, and easier to compare across prompt instances. It can also reduce ambiguity by specifying which fields, labels, or values are expected. However, a schema may constrain the answer too strongly, hide uncertainty, or encourage the model to fill fields even when the input does not support them. The model may also produce malformed structured output or values outside the allowed answer space. The pattern is therefore more reliable when the schema includes instructions for missing, uncertain, or unsupported values.

**Example.** For example, a user may ask the model to classify a bug report using a fixed schema and a constrained set of severity labels.

```
Answer the following question: Classify the bug report.
Use the following output schema: JSON object
Populate the following fields:
- summary: string
- severity: one of ["low", "medium", "high", "critical"]
- category: one of ["bug", "enhancement", "question"]
- rationale: string
Use null when a field cannot be determined from the report.
Bug report: {{bug_report}}
```

**Variants.** Variants include format-only specifications, field schemas, answer-space restrictions, and constrained-vocabulary instructions. In a format-only variant, the prompt specifies the output format, such as CSV, Markdown, XML, or JSON, without defining detailed field constraints. In an answer-space variant, the prompt restricts the response to a predefined set of labels or values. In a constrained-vocabulary variant, the prompt restricts the terms or vocabulary that may appear in the response. Answer-engineering techniques may also include external answer extractors; those extractor mechanisms are outside the scope of this taxonomy unless the constraint is expressed as a textual structure inside the prompt.

**Other names.** Related source concepts merged under this pattern include *Output Format*, *Answer Space*, and *Constrained Vocabulary Prompting*.

# 6 Discussion

We now discuss limitations of our taxonomy and method. We also discuss future perspectives of this taxonomy.

## 6.1 Threats to validity

The taxonomy was constructed through extraction, deduplication, semantic consolidation, scope filtering, categorization, and component annotation. These activities involve interpretive decisions, and several threats to validity should be considered.

The taxonomy is based on five selected sources that organize prompt engineering knowledge through surveys, catalogs, or categorizations. These sources provide broad coverage of prompt engineering concepts, but they do not constitute an exhaustive systematic literature review of all primary studies on prompting. As a result, some prompt patterns may be absent from the taxonomy, especially recently proposed techniques or patterns discussed outside the selected sources.

A central threat concerns the definition of "prompt pattern". Existing work uses related terms such as "prompting technique," "prompting method," and "prompting strategy" in different ways. Our taxonomy adopts a narrower definition as stated in Definition 4. Although it clarifies the scope of the taxonomy, it excludes useful prompting practices such as multi-turn workflows, tool-use protocols, retrieval-augmented generation pipelines, and fine-tuning methods. These exclusions are intentional, but they mean that the taxonomy should not be interpreted as a complete taxonomy of all prompt engineering techniques.

Some patterns may also behave differently across models, domains, or tasks. The taxonomy should therefore be understood as a catalog of reusable prompt structures, not as evidence that each pattern improves output quality in all contexts.

The consolidation of source concepts into canonical patterns required judgments about whether concepts were equivalent, overlapping, variants of a broader pattern, or distinct. For example, concepts with different names may express the same underlying prompting intention, whereas similar names may refer to different structures or levels of granularity. Incorrect merging could hide meaningful distinctions between patterns, while insufficient merging could preserve redundant entries. We mitigated this threat by recording source names, aliases, and variants, and by separating name-level deduplication from semantic consolidation.

Assigning patterns to prompting strategy categories and prompt component annotations also involved subjective decisions. Some patterns can plausibly belong to more than one category or be instantiated in several prompt components. For instance, a pattern may both constrain output and guide reasoning, or may appear in both directive and constraint components. To mitigate this threat, the classification process involved the two authors. One author performed the initial extraction, deduplication, semantic consolidation, scope filtering, taxonomy categorization, and component annotation. The second author validated the resulting decisions by inspecting the source candidates, exclusion decisions, canonical pattern names, category assignments, and component annotations. Unclear cases were discussed and resolved through iteration until both authors agreed on the final taxonomy. The validation focused on four types of decisions: (1) whether a source candidate satisfied the scope of the taxonomy; (2) whether overlapping candidates should be merged into a single canonical pattern; (3) whether a candidate should be represented as a variant or combination of broader patterns; (4) whether the assigned prompting strategy and component annotations were appropriate.

## 6.2 Evolution of patterns and categories

The taxonomy is temporally bounded. Prompt engineering is evolving rapidly, and new prompting practices, model capabilities, and interfaces may lead to new patterns or require revisions to existing categories. The taxonomy is intended to evolve as new prompting practices are reported. New candidates should be assessed against the scope defined in Section 2 and assigned to one of the following outcomes:

1. **Retain as a new pattern**: the candidate is a recurring textual structure, has a distinct prompting intention, and can be instantiated within a single-turn textual prompt.

2. **Record as an alias**: the candidate uses a different name for an existing pattern.

3. **Record as a variant**: the candidate modifies or specializes an existing pattern without changing its main prompting intention.

The category structure should evolve more conservatively than individual pattern entries. A new pattern should first be classified within the existing top-level categories. A new subcategory should be introduced only when several patterns share a recurring prompting intention that is not adequately represented by the current subcategories. A new top-level category should be introduced only when the existing categories cannot accommodate a recurring family of prompt patterns without obscuring their purpose.

Component annotations may be revised independently from the category structure. Because a pattern can be instantiated in more than one prompt component, component annotations should be treated as descriptive metadata rather than fixed membership constraints. Changing a component annotation refines where the pattern can appear in a prompt; it does not necessarily change the identity of the pattern.

Future versions of the taxonomy should document added patterns, aliases, variants, merged entries, exclusions, renamed patterns, revised classifications, and changed component annotations. This versioning is necessary so that studies using the taxonomy can cite the specific catalog version on which their analysis is based.

## 6.3 Relations between patterns

The taxonomy assigns each pattern to one primary prompting strategy, but prompt patterns are not mutually exclusive. A single prompt may instantiate several patterns, and some patterns may overlap in structure, intention, or prompt component. These relations are useful for prompt design, prompt analysis, and future taxonomy maintenance.

We distinguish five practical relation types.

1. **Composition.** Several patterns can be instantiated in the same prompt. For example, a code-review prompt may combine *Persona* to assign the role of a security reviewer, *ContextManager* to restrict the review to security concerns, *SchemaSpecs* to require a structured output, and *SelfCalibration* to ask the model to report confidence. Composition is expected because patterns often operate on different prompt components.

2. **Specialization.** A pattern may specialize another pattern by adding a more specific structure or use condition. For example, a multi-persona prompt can be viewed as a specialization of *Persona*. Similarly, constrained-vocabulary prompting is represented as a specialization of *SchemaSpecs* because it restricts the admissible values or terms in the output.

3. **Combination.** Some reported techniques combine two or more broader patterns. For example, few-shot chain-of-thought combines example-based prompting with explicit reasoning. In such cases, we avoid creating a new canonical pattern unless the combination has a distinct prompting intention and a stable reusable structure.

4. **Boundary overlap.** Some patterns are close in surface form but differ in primary intention. For example, *Persona* and *ContextManager* both influence context, but *Persona* assigns a role or perspective, whereas *ContextManager* controls the contextual information or lens to use. Similarly, *SchemaSpecs* and *Template* both constrain output form, but *SchemaSpecs* defines fields, values, or schemas, whereas *Template* asks the model to preserve and fill a reusable template.

5. **Tension.** Patterns may interact in ways that weaken or constrain each other. For example, strict *SchemaSpecs* may limit the space available for reasoning explanations, while a context-reset variant of *ContextManager* may remove prior role or output instructions. These tensions do not make the patterns incompatible, but they indicate that their order, wording, and scope should be specified carefully.

A study of the relations between patterns should be conducted in the future. Documenting these relations helps make the taxonomy more useful to operationalize prompt patterns, detect them in prompts, or analyze how they are combined in empirical studies.

# 7 Conclusion

Prompt engineering knowledge is increasingly documented, but it remains difficult to reuse when prompt patterns are described with inconsistent terminology, heterogeneous representations, and varying levels of granularity. In this report, we presented a taxonomy of prompt patterns for single-turn, text-based interactions with LLMs. Using a reproducible process, we consolidated concepts from five sources into 30 canonical prompt patterns.

The taxonomy organizes patterns primarily by prompting strategy and annotates them with the prompt component types in which they can be instantiated. This structure supports the description, comparison, and reuse of prompt patterns, particularly in empirical software engineering studies involving LLMs. The accompanying online catalog makes the taxonomy inspectable and provides a basis for future extension as new prompting practices emerge.

The taxonomy is intentionally scope-bounded and does not aim to cover all prompt engineering techniques. It provides a focused catalog of reusable textual structures that can be instantiated within a single prompt. Future work can extend this taxonomy by incorporating new patterns, refining pattern relations, and studying how these patterns are used and combined in real prompts.

## Acknowledgments

This work was funded by the Mitacs Accelerate grant IT38322 partially funded by Draft & Goal `https://dng.ai/`. In particular, we thank Vincent Terrasi, Sharanya Akkenapally, and Nabil Tayeb for their guidance and support.